\documentclass[journal=jacsat,manuscript=article]{achemso}

\usepackage[version=3]{mhchem} 
\usepackage{multirow}
\usepackage{hyphenat}

\author{Gauri Arora}
\altaffiliation{Shared first authorship}
\affiliation[Indian Institute of Technology Delhi]
{Department of Physics, Indian Institute of Technology Delhi, Hauz Khas, New Delhi 110016, India}
\author{Ankit Butola}
\altaffiliation{Shared first authorship}
\email{ankit.butola@uit.no}
\affiliation[UiT The Arctic University of Norway]
{Department of Physics and Technology, UiT The Arctic University of Norway, Tromsø, Norway}
\author{Ruchi Rajput}
\affiliation[Indian Institute of Technology Delhi]
{Department of Physics, Indian Institute of Technology Delhi, Hauz Khas, New Delhi 110016, India}
\author{Rohit Agarwal}
\affiliation[UiT The Arctic University of Norway]
{Department of Computer Science, UiT The Arctic University of Norway, Tromsø, Norway}
\author{Krishna Agarwal}
\affiliation[UiT The Arctic University of Norway]
{Department of Physics, UiT The Arctic University of Norway, Tromsø, Norway}
\author{Alexander Horsch}
\affiliation[UiT The Arctic University of Norway]
{Department of Computer Science, UiT The Arctic University of Norway, Tromsø, Norway}
\author{Dilip K Prasad}
\affiliation[UiT The Arctic University of Norway]
{Department of Computer Science, UiT The Arctic University of Norway, Tromsø, Norway}
\author{Paramasivam Senthilkumaran}
\affiliation[Indian Institute of Technology Delhi]
{Department of Physics, Indian Institute of Technology Delhi, Hauz Khas, New Delhi 110016, India}

\title[An \textsf{achemso} demo]
  {Taxonomy of hybridly polarized Stokes vortex beams}

\abbreviations{IR,NMR,UV}

\begin{document}


\begin{abstract}
 Structured beams carrying topological defects, namely phase and Stokes singularities, have gained extensive interest in numerous areas of optics. The non-separable spin and orbital angular momentum states of hybridly polarized Stokes singular beams provide additional freedom for manipulating optical fields. However, the characterization of hybridly polarized Stokes vortex beams remains challenging owing to the degeneracy associated with the complex polarization structures of these beams. In addition, experimental noise factors such as relative phase, amplitude, and polarization difference together with beam fluctuations add to the perplexity in the identification process. Here, we present a generalized diffraction-based Stokes polarimetry approach assisted with deep learning for efficient identification of Stokes singular beams. A total of 15 classes of beams are considered based on the type of Stokes singularity and their associated mode indices. The resultant total and polarization component intensities of Stokes singular beams after diffraction through a triangular aperture are exploited by the deep neural network to recognize these beams. Our approach presents a classification accuracy of 98.67\% for 15 types of Stokes singular beams that comprise several degenerate cases. The present study illustrates the potential of diffraction of the Stokes singular beam with polarization transformation, modeling of experimental noise factors, and a deep learning framework for characterizing hybridly polarized beams.

\end{abstract}

\section{Introduction}
Phase singularities are associated with the orbital angular momentum (OAM) \cite{COULLET1989403}, and circular polarization is related to spin angular momentum (SAM) of light \cite{Angelsky:12}. The superposition of two beams in orthogonal polarization states with  at least one of the beams carrying phase singularity leads to the formation of Stokes singularities \cite{FREUND2002251, freund2002stokes, RuIJO2020}. Beams with Stokes singularities are called Stokes singular beams (SSBs), where light's spin and orbital angular momentum are coupled together. Several methods  such as interferometric techniques \cite{Arora:19}, implementation of spatial light modulators (SLM) \cite{Wang:07, Gao:19}, stress engineered optics \cite{Ariyawansa:19}, and spatially inhomogeneous wave plates \cite{Radhakrishna:21} have been reported to generate these beams.
The polarization distribution associated with these beams adds an extra degree of freedom in optical communication which increases the channel capacity \cite{Zhao:15, Wang:16, communnew}. The exotic properties of these beams have been exploited in laser beam shaping \cite{Han:11}, super-resolution microscopy \cite{Kozawa:18}, chirality measurement \cite{samlan2018spin}, image processing \cite{BhargavaRam:17}, robust beam engineering\cite{Lochab:17}, Mobius strips generation \cite{FREUND20057} and among others.

The role of SSBs in various applications depends on the net OAM (mode indices) present in these beams. The Stokes index used to define SSBs does not provide complete information of the vector beams in terms of their net OAM content.  Therefore, it is necessary to identify these singularities based on their superposition modes to deploy them for a particular application. However, the detection techniques of these beams are limited to a few, including Stokes polarimetry \cite{GoldCRC12}, interferometric \cite{PhysRevE.65.036602} and diffraction techniques \cite{Gauri_SR2020}. These detection techniques can not deal with all the degenerate states associated with Stokes singular beams. In addition, unavoidable experimental fluctuations in phase, amplitude, and polarization together with other beam fluctuations, makes detection a challenging task. Recently, artificial intelligence (AI) has emerged as a powerful tool to boost scientific research in the field of optics and photonics. The AI-based techniques are also being adopted for the identification of multi-singularity structured field, orbital angular momentum of vortex beams, and among others \cite{Wangmulti},\cite{8891699}. In vector regime, machine learning-based detection of vector vortex beams is presented in \cite{PhysRevLett.124.160401}. However, the classification method adopted here is based on Stokes polarimetry which does not consider the degeneracy present in the SSBs. In addition, the report focuses on the classification of vector vortex beams i.e., a subset of Stokes singularities. Nevertheless, there is no report to date that deals with identifying all the Stokes singularities and their degenerate states.

Here, we present the detection of all Stokes singularities by exploiting the diffraction and polarization transformation patterns of the singular beams assisted with the deep neural network. A deep neural network provides an excellent framework for detecting hybridly polarized beams due to its ability to recognize indiscernible features in the intensity images, that would not be detectable merely by intensity-sensitive measurements. The method employed here is based on the diffraction of singular beams through an equilateral triangular aperture in combination with polarization transformation. The resultant total and component intensities of a particular Stokes singular beam after diffraction are utilized for classification purposes. A total of 15 classes of beams are first simulated and further experimentally generated based on the type of Stokes singularity and the associated mode indices.

Next, the classification of all Stokes singular beams is performed and compared by using five different deep neural networks. These networks are trained with mixed simulation and experimental datasets which consist of 15300 images (15000 simulated and 300 experimentally acquired). A total of 90\% of the simulated images and 50\% of the experimentally acquired images are used for training and validation purposes. Finally, the optimized trained network is tested on 10\% of the simulated image and 50\% of the experimental datasets. Separate testing accuracy for both simulated and experimental datasets is shown to demonstrate the advantage of the current framework. We found ResNet-18 offers the best testing accuracy i.e., 97.67\% and 98.67\% for simulated and experimental data, respectively. Furthermore, the current experimental and computational approach addresses both detection and classification tasks thus, enabling the recognition of total and component intensities of the Stokes singular beams which could be relevant in applications such as super-resolution optical microscopy and optical communication. Our findings manifest the soundness of proposed method for the detection of all Stokes singularities and novel identification method which paves the way to scalable microscopy applications through Stokes beams illumination. 

\begin{figure*}[!ht]
\centering
\includegraphics[width=15cm, height=5.0cm]{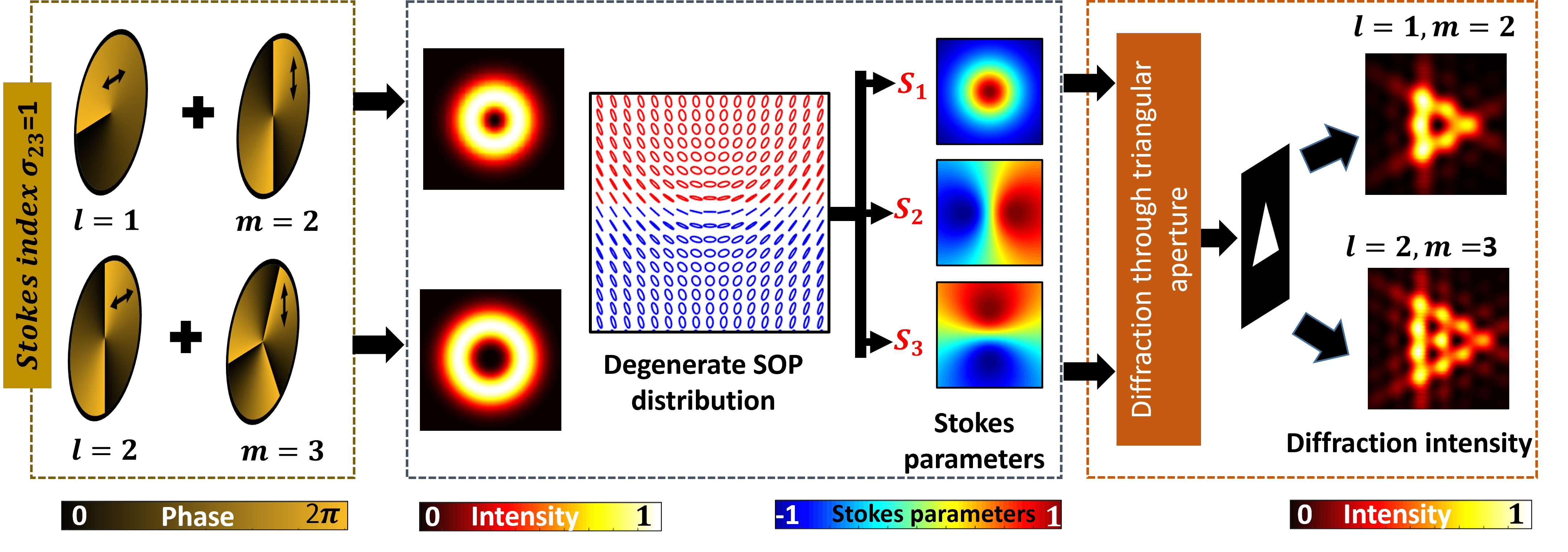}
\caption{{\textbf{An example of polarization degeneracy in Stokes singular beams with Stokes index $\sigma_{23}=1$.} Left: Phase distribution of the two superposing vortex beams characterized by topological charges $l$ and $m$ respectively. Orthogonal polarization basis states are marked in the inset. Centre: Corresponding total intensity, degenerate polarization distribution and Stokes parameter distributions. Right: Respective diffraction patterns produced by the two beams after diffraction through a triangular aperture.}}
\label{Fig1}
\end{figure*}


\section{Methods} \label{sec:principle}

\subsection{Stokes degenerate states} 
Any state of polarization of light can be defined using four Stokes parameters $S_0$, $S_1$, $S_2$, and $S_3$ \cite{GoldCRC12}
\begin{equation}
\begin{split}
S_0 &= |E_x|^2+|E_y|^2; \quad    
S_1 = |E_x|^2-|E_y|^2;\\
S_2 &= 2Re(E_x^*E_y); \quad
S_3 = 2Im(E_x^*E_y).	
\end{split}
\label{eq:SP}
\end{equation}
For inhomogeneously polarized beams, Stokes parameters are functions of $x$ and $y$. Here, the Stokes parameter $S_0$ gives total intensity and other Stokes parameters give differences in intensities of orthogonal polarization components. Using these Stokes parameters, Stokes fields $S_{jk}=S_j+iS_k$ can be constructed which can be $S_{12}$, $S_{23}$, and $S_{31}$. The corresponding Stokes phases are $\phi_{12}$, $\phi_{23}$, and $\phi_{31}$ where $\phi_{jk}$ is argument of Stokes field defined as: $\phi_{jk}=\tan^{-1} (S_k/S_j)$. The Stokes singularities are points of phase singularities in these Stokes phases.
 Stokes singularities $\phi_{12}$, $\phi_{23}$, and $\phi_{31}$ provide information about the phase singularities present in orthogonal polarization basis states, namely,  circular (R, L), linear (H, V), and linear (D, A) bases respectively. Here, R, L, H, V, D, and A represent right circular, left circular, horizontal, vertical, diagonal, and anti-diagonal polarization states. Various types of Stokes singularities are reported in the literature which includes point and line singularities in 2D transverse plane of the beam, and singularities present in 3D fields which include Mobius strips, links and knots \cite{FREUND20057}. In this article, we deal with point Stokes singularities present in 2D transverse plane of the beam.

At $\phi_{12}$ Stokes singular point, polarization azimuth is indeterminate; hence, these singularities are also known as polarization singularities. The other two types of Stokes singularities are called Poincar\'{e} vortices. Under paraxial approximations, Stokes singular beams can be mathematically expressed as the superposition of two beams in orthogonal polarization states such that at least one of them carries phase singularity and is given by
\begin{multline}
 {\left|{\psi_{l,m}(r,\theta)}\right\rangle}= \cos(\chi+\frac{\pi}{4})\left|\psi_{P}^l(r, \theta)\right\rangle + \sin(\chi+\frac{\pi}{4})
 \\ \exp(i2\gamma) \left|\psi_{Q}^m(r, \theta)\right\rangle
\end{multline}
where
\begin{equation}
  \left|\psi_{P}^l(r, \theta)\right\rangle = \psi^l(r)\exp(il\theta)\hat{P}
\end{equation}
\begin{equation}
  \left|\psi_{Q}^m(r, \theta)\right\rangle = \psi^m(r)\exp(im\theta)\hat{Q}.
\end{equation}
Here, $2\chi$ and $2\gamma$ decide the weighting factor and phase difference between the two superposing beams respectively. The variables $l$ and $m$ represent topological charges of the vortex beams used to construct SSBs, and $\hat{P}$, $\hat{Q}$ represent orthogonal polarization basis states. The ${\left|{\psi_{l,m}(r,\theta)}\right\rangle}$ gives the resultant amplitude distribution due to the superposition of two beams. Bright Stokes singularities are formed when one of the superposing beams is a Gaussian beam. The superposition of phase singular beams in orthogonal polarization states result in dark Stokes singularities. These singularities can be identified using a Stokes index $\sigma_{jk}$ which is defined as $\sigma_{jk}= \frac{1}{2\pi} \oint \nabla \phi_{jk} \cdot dl$ where $\phi_{jk}$ represents the Stokes phase, and $dl$ is the closed path of integration around the singular point. Stokes index can also be defined as $\sigma_{jk}=m-l$, where $m$ and $l$ are topological charges of component phase singular beams respectively.

Stokes singular beams in which the singularities are present at the center of the beam are represented as points on the hybrid-order Poincar\'{e} spheres(HyOPS) or higher-order Poincar\'{e} spheres(HOPS). These spheres are geometrical constructions where one or both the poles of the spheres represent phase singular beams in orthogonal polarization states. Stokes singular beams with the same Stokes index can have different intensity, polarization, and Stokes phases distributions, which result in degenerate Stokes index states\cite{Gauri_SR2020}. Stokes singular beams generated using the superposition of phase singularities with topological charges $l$ and $m$ such that $m-l=constant$, and $l\neq0, m\neq0$ are polarization degenerate. All the beams represented on a surface of a particular HyOPS/HOPS are Stokes index degenerate \cite{Gauri_SR2020}. Also, the SSBs on a particular longitude of a hybrid order Poincar\'{e} sphere have similar polarization distribution but different polarization gradients \cite{Arora_2021, Gauri_SR2020}.

The degenerate polarization singular beams are composed of either different mode indices ($l,m$) or same mode indices but with different relative weightage of $l$ and $m$ and therefore contain different optical properties. For example, a partially coherent beam is shown to produce different depolarization effects despite of having the same Stokes index and polarization distribution \cite{Guo:16}.  Further, polarization degenerate beams cannot be differentiated using the Stokes polarimetry method because of same polarization distribution. Hence, the identification of SSBs based on their mode indices is important.
Figure \ref{Fig1} shows one example where the Stokes singularities of a given index are composed of different mode indices but are polarization degenerate. The component phases and the corresponding polarization states (marked on the phase maps) from which the SSBs are composed of are shown in the left side of the figure. The corresponding resultant intensity distributions, degenerate polarization distribution and the respective Stokes parameter distributions ($S_1, S_2, S_3$) are shown in the middle of the figure. Due to the same polarization distributions, both the beams have same Stokes parameter distributions. On the right side of the figure, the respective diffraction intensity distributions produced by two distinct SSBs are shown to distinguish the degenerate polarization states. 

\begin{figure*}[!ht]
    \centering
  \includegraphics[trim=26mm 51mm 56mm 44mm, clip, width=16cm, height=6cm]{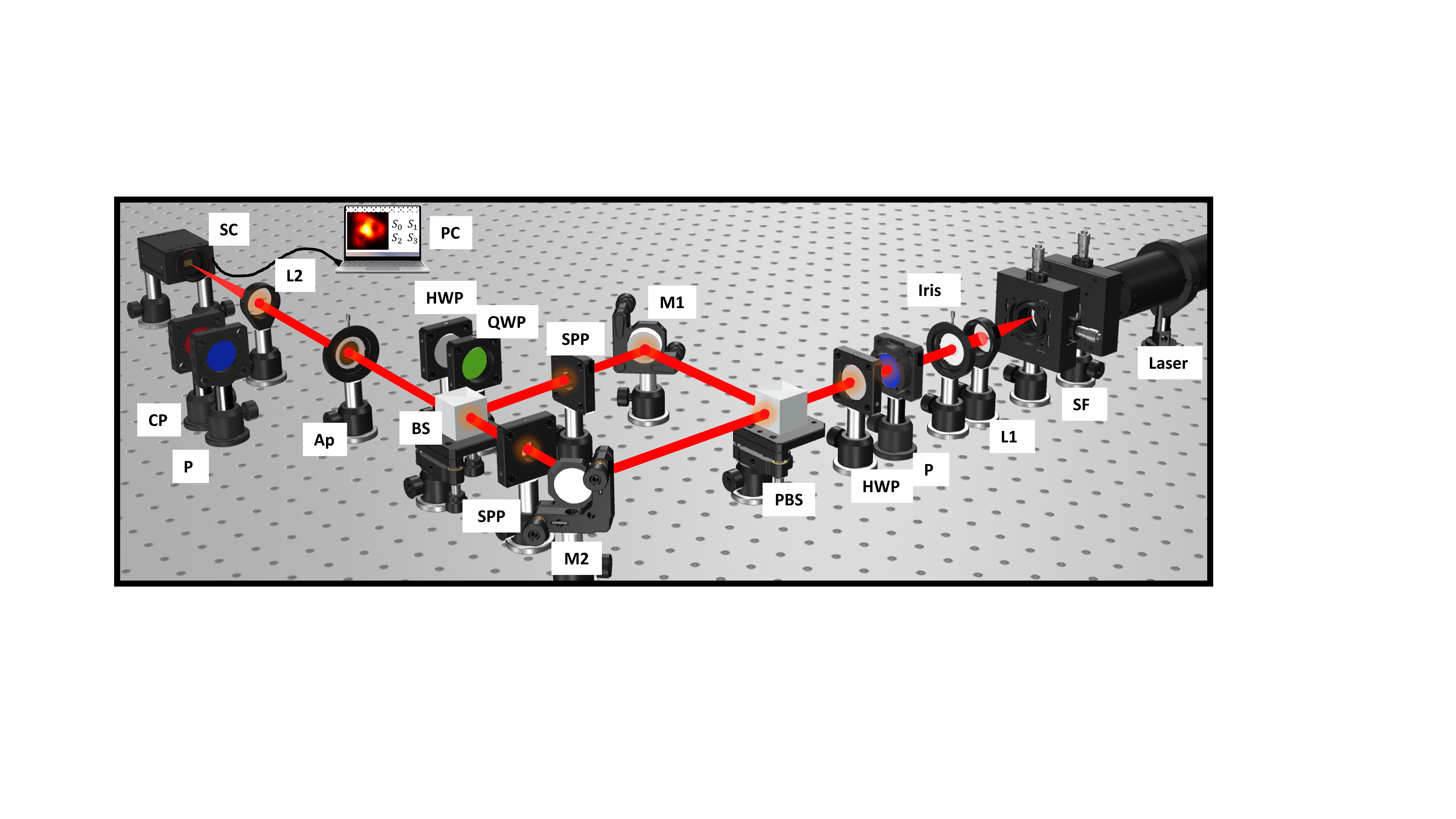}
    \caption{\textbf{Experimental set up for generation and detection of Stokes singular beams}. SF: Spatial filter assembly, L1, L2: Lenses, Iris: amplitude aperture to control the size of the beam, P: Polarizer,  H(Q)WP: half(quarter) wave plate, (P)BS: (polarizing) beam splitter, M1, M2: Mirrors, SPP: spiral phase plate, Ap:  Triangular aperture, CP: Circular polarizer,   SC: Stokes camera, PC: Personal computer.}
    \label{Fig2}
\end{figure*}
\begin{figure*}[!ht]
    \centering
    \includegraphics[width=\textwidth, trim={4mm 4mm 5mm 2mm}, clip]{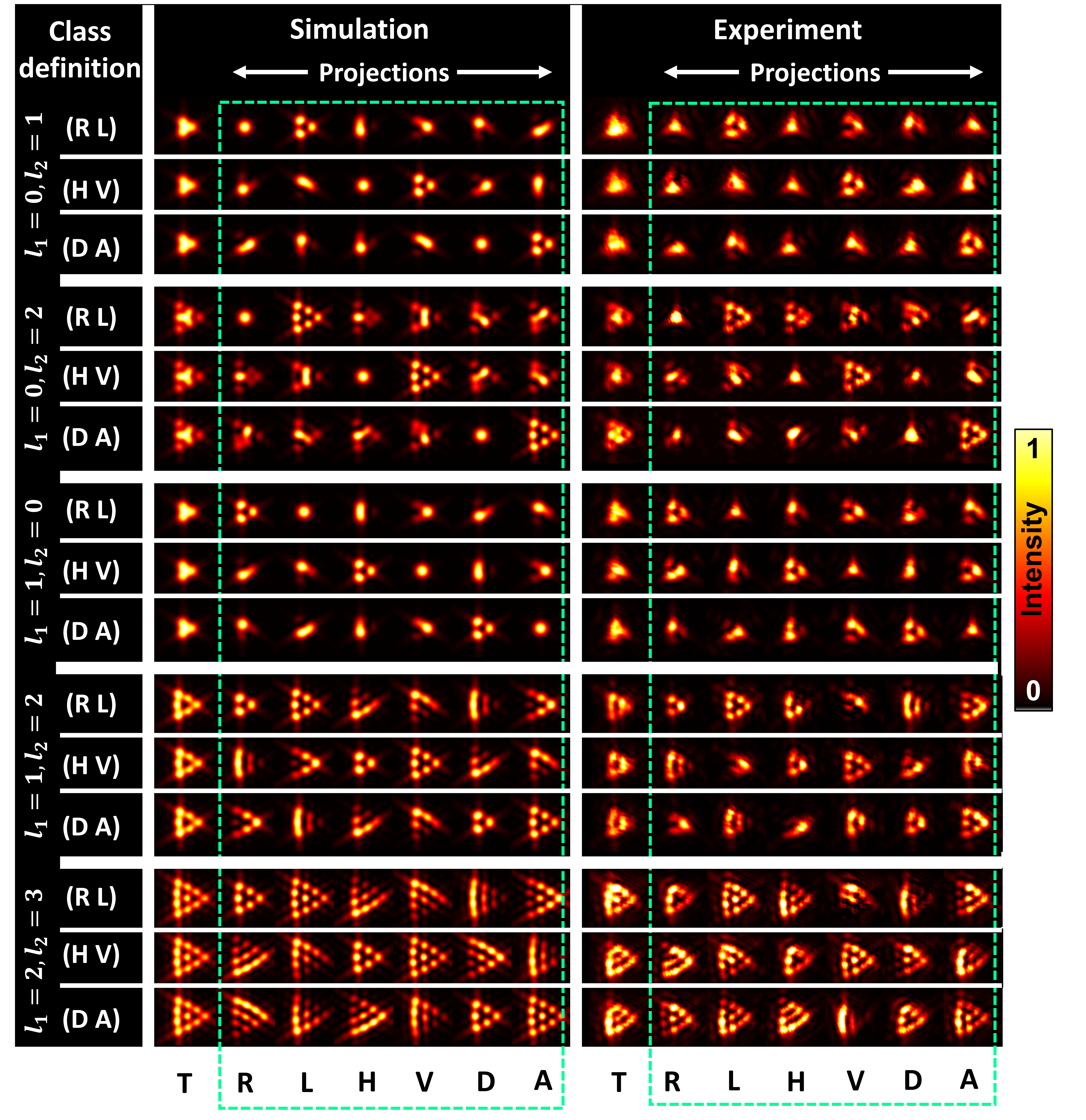}
    \caption{\textbf{Simulation and experimental results of total diffraction intensity patterns (T) and corresponding polarization projections (R, L, H, V, D, A) for various Stokes singular beams.} For each mode indices $(l=l_1,m=l_2)$, superposition in three different polarization basis (R,L; H,V; D,A) is considered.}
    \label{Fig3}
\end{figure*}
\begin{figure*}[t]
    \begin{center}
    \includegraphics[width=\textwidth, trim={0.0cm 12.7cm 6.3cm 0.0cm}, clip]
    {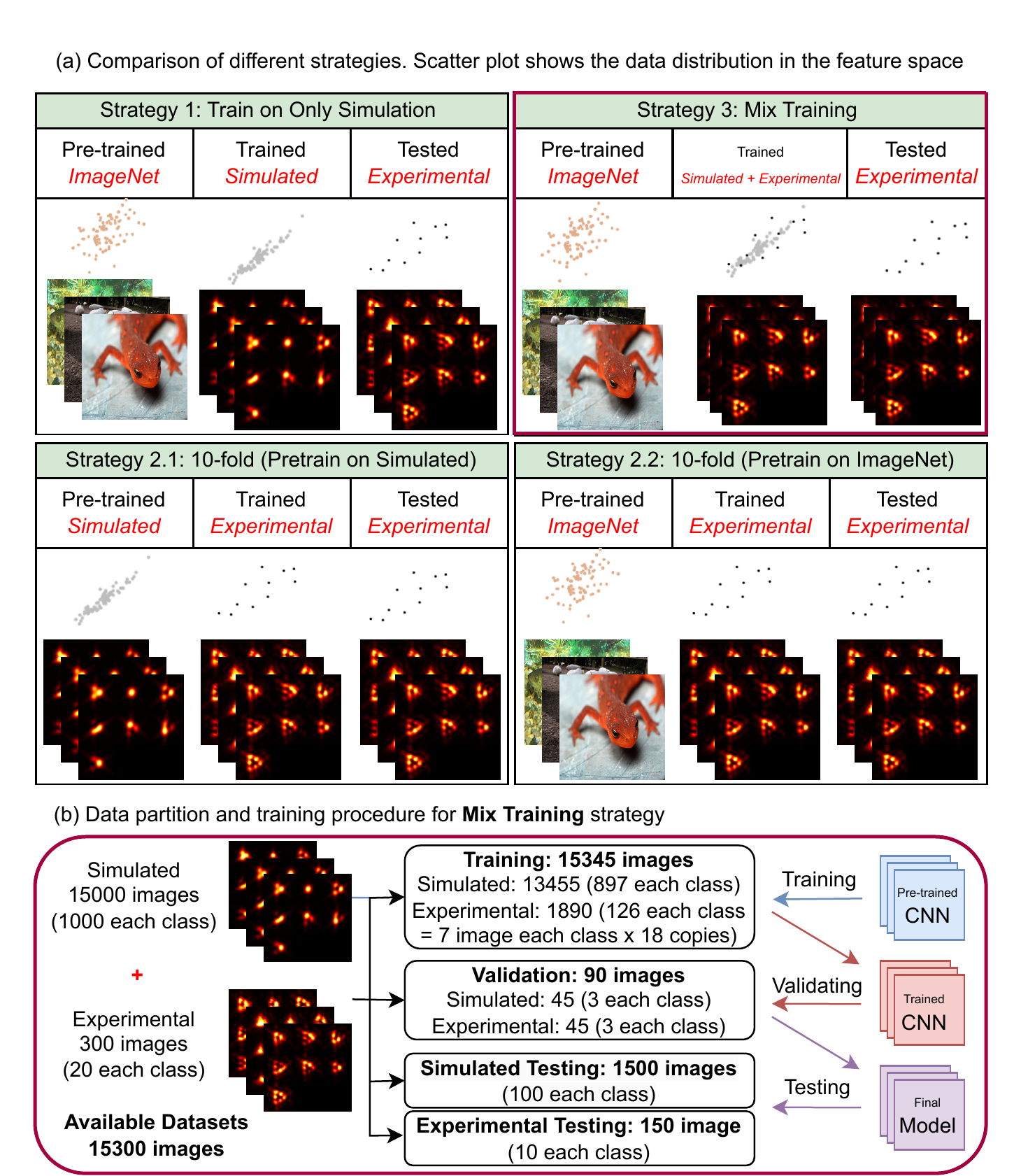}
    \end{center}
    \caption{(a) Comparison of different training strategies: We compare the data distribution of the 3 different training strategies. The scatter plot shown here is for illustration of data distribution. (b) Data partition for training, validation and testing of the neural network under mix training strategy}
    \label{fig:dataprep}
\end{figure*}

\subsection{Detection of Stokes singularities}

 The diffraction of Stokes singular beams through an equilateral triangular aperture lifts the degeneracy associated with mode indices (refer to Fig. \ref{Fig1}). However, the SSBs generated using same combination of mode indices irrespective of their orthogonal polarization states results in diffraction degenerate states. Thus, to completely lift the degeneracy associated with SSBs, polarization transformation after diffraction is obligatory. 

The transmittance function of an equilateral triangular aperture with side length $a$ can be mathematically written as \cite{Gauri_SR2020}:
 \begin{equation}
 \begin{split}
T(x,y) &=  1 \quad  \text{for} \quad \frac{-a}{2}\leq  x\leq \frac{a}{2}\quad\text{and}\quad
    0 \leq y \leq -\sqrt{3}(|x|-\frac{a}{2})\\
 &= 0 \quad \text{otherwise}
 \end{split}
 \end{equation}

 The field $\left|\psi_{l,m} (r, \theta)\right\rangle$ in cartesian coordinate system is written as $\left|\psi_{l,m} (x, y)\right\rangle$ where $x=r\cos\theta$ and $y=r\sin\theta$. The field just after the aperture with aperture function $T(x,y)$ is given by:
 \begin{equation}
    {\left|{E_{T}}\right\rangle}=T(x,y){\left|{\psi_{l,m} (x, y)}\right\rangle}
 \end{equation}
 The far-field diffraction pattern is produced at the back focal plane of a lens with focal length $f$ and is given by:
 \begin{equation}
    {\left|{\mathcal{F}}(u,v)\right\rangle}=\frac{B}{i\lambda f}
    \int_{-\infty}^{\infty} \int_{-\infty}^{\infty} T(x,y){\left|{\psi_{l,m}}\right\rangle}e^{-i\frac{2\pi}{\lambda f}(ux+vy)}dxdy
 \end{equation}
 where $(u,v)$ denotes the coordinates of the Fourier plane, $B$ is a constant and $\lambda$ defines the wavelength of light. The diffraction of the individual superposing beams is calculated independently and added together to obtain the total diffraction pattern of a Stokes singular beam. Notably, the diffraction of these beams can be carried out using other apertures too, that will result in different diffraction intensity patterns depending on their shape and symmetry. However, the aperture should be chosen such that it can differentiate between positive and negative phase singularities with same magnitude of topological charge.\\
 
 Diffraction is a wave phenomenon that can be used to identify the OAM of light. Since SAM and OAM are coupled in SSBs, diffraction and Stokes polarimetry techniques alone cannot distinguish between distinct SSBs. Therefore, here we report the
detection of these beams by combining diffraction based Stokes polarimetry method with the deep learning techniques. The ability of deep learning-based models to recognize intricate features in intensity images makes it an excellent tool to classify Stokes singular beams. 
Respective total diffraction intensity distributions and all the six polarization projections (R, L, H, V, D and A) are utilized as a set to serve as input channels for training the deep neural network (DNN). Of the infinitely many possible classes of SSBs, we selected a total of 15 classes of SSBs. These classes include: $\phi_{jk}^{01}$, $\phi_{jk}^{02}$, $\phi_{jk}^{10}$, $\phi_{jk}^{12}$, $\phi_{jk}^{23}$ where ${j,k}$ run cyclically from $1-3$ and $j\neq k$. Here, in $\phi_{jk}^{lm}$, indices $j$ and $k$ decide the orthogonal polarization states, and, $l$ and $m$ represent the topological charges of the component vortex states of the Stokes singular beam where the Stokes singularity is present at the center. In each of the superpositions, other types of Stokes vortices are also present at different locations \cite{Arora:19}. The corresponding mode indices for these classes are given by (0,1), (0,2), (1,0), (1,2), and (2,3). Among each mode indices, three pairs of polarization basis (R, L; H, V; D, A) are considered, contributing to 15 classes.  

\section{Experimental setup}
The experimental setup used to generate and detect SSBs is depicted in Fig. \ref{Fig2}. The light from the He-Ne laser (632nm) is spatially filtered and collimated using lens L1. A $45^\circ$ polarized beam is generated using a combination of polarizer (P) and half wave plate (HWP). The beam is launched into a modified Mach Zehnder type interferometer where two orthogonally linearly polarized beams in two arms of the interferometer acquire different topological charges using spiral phase plate (SPP). For generating bright Stokes singularity, one SPP was removed from one of the arms of the interferometer. The two beams combine after the beam splitter (BS). The quarter wave plate (QWP) after BS is used to change linear basis superposition to circular basis superposition and the HWP after the BS is used to change (H,V) linear basis to (D,A) linear basis superposition. Further, the beam is allowed to diffract through an equilateral triangular aperture, and the far-field diffraction intensity pattern is recorded using the Stokes camera. The combination of polarizer and QWP after the triangular aperture is used to extract the component intensities of the beam which includes horizontal (H), vertical (V), diagonal (D), anti-diagonal (A), right circular (R) and left circular (L) polarization intensities. The total and polarization components intensities are recorded at three different z-planes for a particular Stokes singular beam. Based on the generation of a particular Stokes index beam, the charge of the spiral phase plate in the two arms of the interferometer is varied.

\section{Results and Discussion}
\subsection{Total diffraction pattern and corresponding polarization projections of Stokes singular beams}
Simulation and experimental results which include total (T) and component diffracted intensities (R, L, H, V, D, A) for different SSBs are depicted in Fig. 3. These component intensities are projections of the resultant beam onto various polarization states. The topological charges of the superposing phase singular beams that form the Stokes singularity are mentioned in the left side of the figure in each case. Each set contains three rows for superposition in three different polarization basis namely circular (R, L), linear (H, V), and diagonal bases (D, A) from top to bottom, respectively.  All the polarization component intensities (R, L, H, V, D, and A) are extracted for each basis of superposition (R,L; H,V; and D,A). Stokes singular beams composed of different mode indices produce different diffraction patterns.
However, SSBs composed of the same mode indices results in the same resultant diffraction pattern irrespective of the polarization associated with component beams, and hence are diffraction degenerate (refer to Fig. \ref{Fig3}). To lift the diffraction degeneracy, polarization transformation is utilized.

In the next section, we describe the deep learning procedure that is used for the recognition of SSBs. It includes dataset preparation, selection and optimization of deep neural networks followed by the classification results. 

\subsection{Deep Learning Models and Strategy}
\label{ml}

\paragraph{Simulation and Experimental datasets:} The datasets used in the present study consist of both simulated and experimentally acquired images. Possible experimental fluctuations in amplitude, phase, polarization, beam shift, and aperture shape are taken into account while creating simulation data for the identification of SSBs. These fluctuations may affect the accuracy of the detection techniques. Therefore, it is imperative to include these parameters while simulating the datasets for different classes. Moreover, this also helps in the robust training of deep learning networks. In addition, since deep learning models require a large amount of data for training \cite{alom2019state}, robust simulation aids in the quick convergence of the network. Heuristically, we simulate 1000 images for each class, resulting in 15000 total images. The experimental data contained a total of 300 images i.e., 20 images for each class. The experimental datasets were acquired under the specifications mentioned in section 3. 

\paragraph{Comparison Models} We considered five different convolutional neural networks (CNN) based deep learning  models, namely, SqueezeNet \cite{iandola2016squeezenet}, VGG \cite{simonyan2014very}, AlexNet \cite{krizhevsky2017imagenet}, DenseNet \cite{huang2017densely} and ResNet-18 \cite{he2016deep} to apply to our datasets. All of these models are pre-trained on ImageNet data. More information about these models is provided in \textit{Supplementary S1}. The cross-entropy loss is employed and a stochastic gradient descent optimizer is used for back-propagation. Each of the $707 \times 101$ image as shown in Fig. \ref{Fig3} is transformed into a square image of dimension 303$\times$303 by stacking two voids of dimension 101$\times$101 and served as an input image to the neural network. Examples of resultant input image are depicted in Figure \ref{fig:dataprep}. This peculiar dimension is chosen in order to preserve the information by symmetry, since the deep learning models resize the input images to 224$\times$224.

\begin{figure}[!ht]
    \begin{center}
    \includegraphics[width=\textwidth, trim={0.9cm 11.67cm 0.2cm 1.78cm}, clip]
    {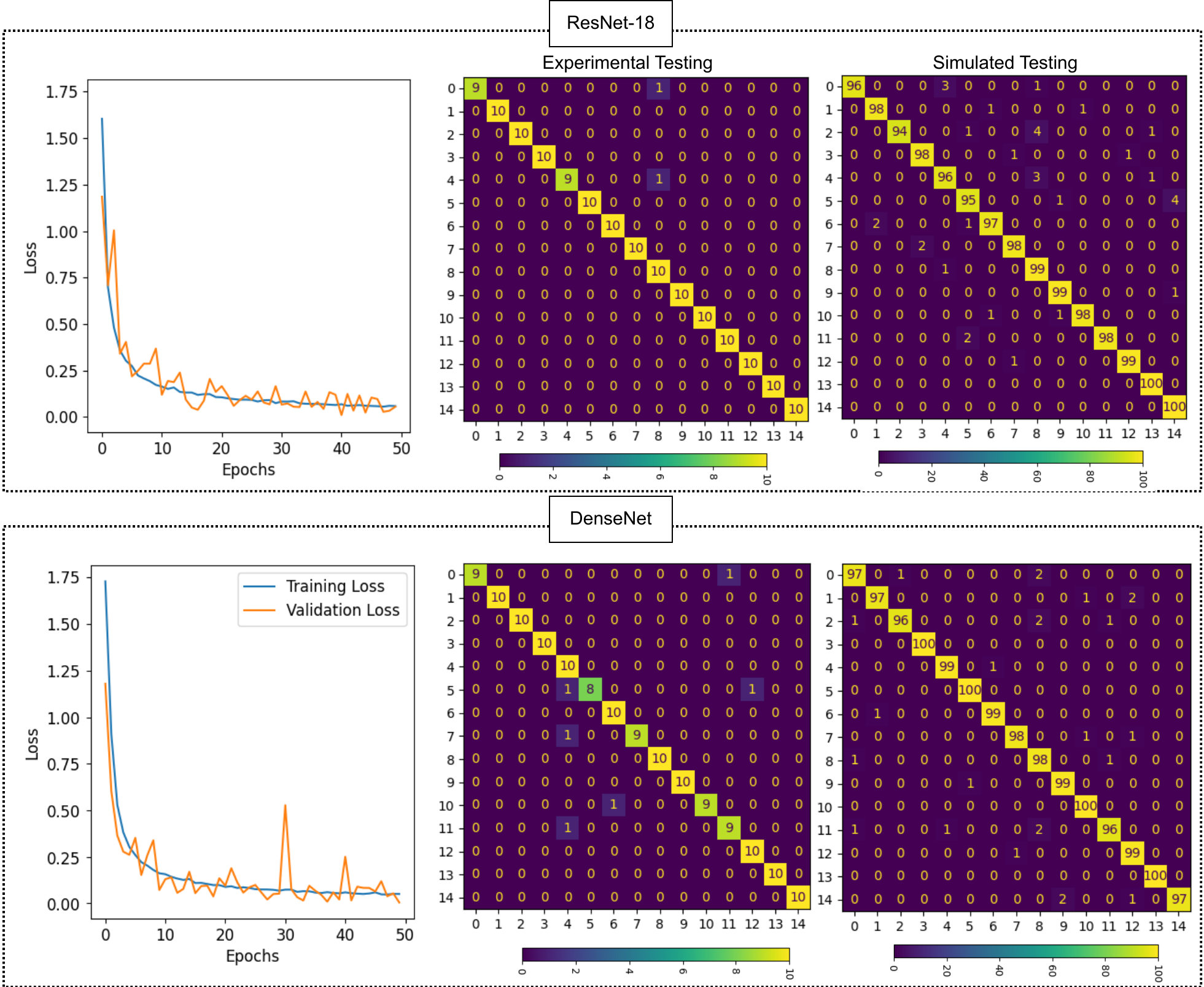}
    \end{center}
    \caption{\textbf{Loss curve and confusion matrices of ResNet\hyp{}18 and DenseNet}. The loss curve is shown after each epoch on the training and validation dataset. The confusion matrix shows the classification accuracy of the network on experimental and simulated testing datasets. Class representation: 0: $\phi_{12}^{01}$, 1: $\phi_{12}^{02}$, 2: $\phi_{12}^{10}$, 3: $\phi_{12}^{12}$, 4: $\phi_{12}^{23}$, 5: $\phi_{23}^{01}$, 6: $\phi_{23}^{02}$, 7: $\phi_{23}^{10}$, 8: $\phi_{23}^{12}$, 9: $\phi_{23}^{23}$, 10: $\phi_{31}^{01}$, 11: $\phi_{31}^{02}$, 12: $\phi_{31}^{10}$, 13: $\phi_{31}^{12}$, 14: $\phi_{31}^{23}$. Check the labelling of classes here and in text.}
    \label{fig:loss}
\end{figure}

\paragraph{Training strategy} We employed 3 different strategies to train and test our model as shown in Fig. 4. Figure 4 depicts the different strategies used for training and testing to achieve the best classification accuracy. A scatter plot is shown in Fig. 4 to show the data distribution in the feature space. Next, we explain all three training and testing strategies:

(1) \underline{Training with only simulation data}: We trained and validated all five deep-learning models using simulated data and tested on the experimental data. However, this strategy provides very poor classification performance i.e., 43.6\%. Note that this is the best achievable performance  among five deep learning models and is obtained through Alexnet. This is attributed to the fact that experimental data have more variability between the images in each class which is illustrated in the scatter plot in Fig. 4. In addition, slight mismatch between the simulated and experimentally acquired datasets also affects the classification accuracy. This strategy is further discussed in \textit{Supplementary S2}.


(2) \underline{10-fold strategy on experimental data}: We employed a 10-fold strategy \cite{wong2019reliable} to train a ResNet-18 model on experimental data as shown in Fig. 4. ResNet-18 model was first pre-trained on the simulated data followed by fine-tuning using experimental data by a 10-fold strategy (strategy 2.1 in Fig. 4). This strategy works well and provides an accuracy of 93\%. On the other hand, fine-tuning of the ResNet-18 model (pre-trained on ImageNet) on experimental data with a 10-fold strategy is also performed (strategy 2.2 in Fig. 4). Fine-tuning with only experimental datasets provides an accuracy of 92\%. Strategy 2 does not exploit the benefit of simulated data as strategy 2.1 (shown in Fig. 3) improves the final performance by only 1\%. We discuss strategy 2 in more detail in \textit{Supplementary 3}. 

(3) \underline{Mix Training}: Finally, we train the deep-learning model on mix simulated and experimental data \cite{giordani2020machine} for generalized and robust training. Mix training provides the best accuracy of 98.67\% using the ResNet-18 model. The mix strategy is discussed in detail in the next paragraph.

\paragraph{Mix Training:} For robust training of the neural networks, we performed mixed training with simulated and experimental datasets. The dataset is divided into three parts: training, validation and the testing set. The testing set is further categorized into two subparts: simulated testing and experimental testing. A total of five different deep learning models are trained separately and optimized for a fixed batch size of 8. The models are trained on the training set, the best hyperparameters are chosen on the validation set and the performance of each network is shown on the testing set in Table 1. The division of the dataset is shown in Figure \ref{fig:dataprep}(b). A total of 35\% (105 images), 15\% (45 images) and 50\% (150 images) of the experimental images are used to training, validation and testing respectively. From simulated data, 89.7\% (13455 images), 0.3\% (45 images) and 10\% (1500 images) images are used for training, validation and testing purpose respectively. We chose only 0.3\% of simulated images for validation to ensure that both experimental and simulated datasets contained an equal number of images in the validation set. Moreover, the ratio of simulated and experimental images in the training datasets is optimized since strategy 1 confirms that training model with only simulated images offer poor accuracy on experimentally acquired images. To balance the training datasets, all experimental images of the training set (105 images) are copied 18 times resulting 1890 experimental images. This results in ~7:1 ratio of simulated and experimental images in the  training set. This strategy ensures the availability of one experimental image in each batch of 8 images during training of the network. All the models are trained for 50 epochs. The learning rate is optimized for all models and found to be $0.001$ on the validation set. The accuracy is calculated as the percentage of instances rightly classified. The accuracy of each model in all the parts of the dataset is given in Table \ref{table:resultsml}.

\subsection{Labelling hybridly polarized beams and assessing the winning models}
Out of all models used in the present study, ResNet-18 and DenseNet offer the best training accuracy of 97.81$\%$ and 98.36$\%$ respectively (see Table \ref{table:resultsml}). DenseNet performs better in the validation set followed by ResNet-18. DenseNet offers 100$\%$ of validation accuracy while ResNet-18 provides a validation accuracy of 97.78$\%$. In the simulation testing dataset, ResNet-18 and DenseNet give classification accuracy of 98.33$\%$ and 97.67$\%$ respectively. Other architecture i.e., SqueezeNet, VGG and AlexNet offer accuracy of 85.80 $\%$, 95.73$\%$ and 95.00$\%$ respectively. Interestingly, ResNet-18 outperforms all the models in the experimental testing dataset. The classification accuracy on the experimental images using ResNet-18 is found to be 98.67$\%$.

Further to quantify the best-performed model i.e., ResNet-18 and DenseNet, we show the loss curve in Fig. \ref{fig:loss}. Both training and validation  loss is calculated and shown after each epoch here. It is evident from the training loss curve that both models are converging after 50 epochs. Further, confusion matrices of ResNet-18 and DenseNet shown in Fig. \ref{fig:loss} demonstrate the visual clues of the network prediction. Confusion matrices are shown for both the simulation and experimental testing datasets. Diagonal elements of the confusion matrix show correct prediction while off-diagonal elements are the ones wrongly classified by the network. For example, only two images are wrongly classified by ResNet-18 in the experimental testing datasets i.e., one instance of the class $\phi_{12}^{01}$ which is represented as class 0 and class $\phi_{12}^{12}$ which is shown as class 4 in Fig. \ref{fig:loss}. Wrongly classified image of $\phi_{12}^{01}$ and $\phi_{12}^{12}$ are classified as $\phi_{23}^{12}$ (represented as class 8) by ResNet-18. On the contrary, DenseNet wrongly classifies 6 images, and hence performs slightly poorer than ResNet-18. Nonetheless, the experimental classification accuracy of ResNet-18 and DenseNet are 98.33$\%$ and 97.67$\%$ respectively manifesting the benefits of using a deep learning approach to quantify SSBs in their relevant classes. The efficacy of ResNet-18 can be attributed to their residual connections which alleviate the degrading performance due to the vanishing gradient problem in deep neural networks.

A deep learning model requires a substantial amount of data for training purposes. While acquiring a large amount of experimental data is challenging as it requires considerable time and effort, it is imperative to simulate images that exhibit properties similar to experimental images. However, it is almost impossible to include all experimental variations in the simulated images and therefore, our proposed mix training strategy of a combination of simulation and a few experimental images offers a unique solution to overcome the aforementioned issue and successfully leverage the performance of CNNs for the classification of Stokes singular beams. Moreover, the superior performance by ResNet-18 asserts that a simpler network with a skip connection performs better in the case of Stokes singular beams.

Taken together, our work presents important steps towards understanding Stokes singular beams by a unique experimental approach and should help to democratize their applications in super-resolution imaging, optical communication, beam shaping and potentially new label-free imaging technique beyond what can be achieved in the present label-free optical microscopy community. Our ability to identify the specific class of hybridly polarized beam can be further upscaled to real-time identification of vortex beams which can act as a bridge between beam shaping and microscopy community to understand and choose the appropriate beams for their precise applications.


\setlength{\tabcolsep}{4.5pt}
\begin{table}[t]
\begin{center}
\caption{Model comparison based on the accuracy is given here. The accuracy is rounded to 2 digits. The \textbf{best} model and performance is highlighted in bold and the \textit{second best} in italics. Sim Testing and Exp Testing represent simulation testing and experimental testing datasets respectively.}

\vspace{2mm}
\label{table:resultsml}
{%
\begin{tabular}{|l|c|c|c|c|}
\hline

  \hline

  \hline
         \multirow{2}{*}{\textbf{Models}} & \multicolumn{4}{c|}{\textbf{Accuracy in (\%)}} \\
         \cline{2-5}
         
& Training & Validation & Sim Testing &  Exp Testing\\
\hline
SqueezeNet \cite{iandola2016squeezenet} & 90.72 & 91.11 & 85.80 & 88.00\\
\hline
VGG \cite{simonyan2014very} & 95.85 & 97.78 & 95.73 & 93.33\\
\hline
AlexNet \cite{krizhevsky2017imagenet} & 94.71 & 92.22 & 95.00 & 94.00\\
\hline
\textit{DenseNet} \cite{huang2017densely} & \textbf{98.36} & \textbf{100.00} & \textbf{98.33} & \textit{96.00}\\
\hline
\textbf{ResNet-18} \cite{he2016deep} & \textit{97.81} & \textit{97.78} & \textit{97.67} & \textbf{98.67}\\
\hline
\end{tabular}%
}
\end{center}
\end{table}
\setlength{\tabcolsep}{4.5pt}




\section{Conclusion}
We demonstrated a generalized experimental and computational framework to study the taxonomy of hybridly polarized beams. The diffraction pattern of Stokes singular beams through an equilateral triangular aperture  and their polarization projections are utilized for training a deep neural network. 
After appropriate training, deep neural networks offer excellent accuracy to classify and detect the structured beams carrying Stokes singularities. The scheme discussed in the article can lift all the degeneracy associated with SSBs. Further, the detection scheme is intensity-based, and mixed simulation and experimental images are exploited to train the neural network, which reduces the requirement of acquiring thousands of experimental datasets. We have achieved 97.67$\%$ and 98.33$\%$ accuracy in unknown simulated and experimental testing datasets, respectively. In addition to the characterization of Stokes singular beams, the present approach can also be applied to other applications such as optical microscopy and optical communication. It can be used to design different label-free techniques with improved resolution. 

\section*{Acknowledgement}
KA and AB acknowledges the European Research Council Starting grant (id 804233). 

\section*{Disclosures}
Authors declare no competing interest.

\section*{Author contribution}
AB, PS, KA conceived the idea and supervised the work. GA, RR, PS designed simulation,  experimental system and planned the experiments. AB, RA, KA provide inputs with simulation study. RA, AB, DKP optimized the computational part and classify the datasets. GA, AB, RR, RA analysed the result and prepared the figure. GA mainly wrote the manuscript and all author contributed revised the manuscript.

\bibliography{achemso-demo}

\end{document}